\documentclass[a4paper,12pt]{article}
\usepackage[utf8]{inputenc}
\usepackage[english]{babel}
\usepackage{hyperref}

\title{On the behavior of orbits of Vanhaecke system on integral surfaces}
\author{Wang Shiwei, M.D. Malykh, L.A. Sevastianov, A.V. Zorin}
\usepackage[centertags]{amsmath}
\usepackage{amsfonts}
\usepackage{amssymb}
\usepackage{amsthm}
\usepackage{graphicx}

\theoremstyle{remark}

\newtheorem*{acknowledgments}{Acknowledgments}
\theoremstyle{definition}

\newtheorem{theorem}{Theorem}

\def\vec{\mathfrak }

\def\Al{\operatorname{Al}}

\begin{document}\large\maketitle

\begin{abstract}
In the 1990s, P. Vanhecke described a Hamiltonian system with two degrees of freedom and a polynomial Hamiltonian integrable in Abelian functions of two variables. This system provides a convenient example of an integrable system in which integral curves are wound on a two-dimensional manifold, an algebraic surface in a 4-dimensional phase space. In this report, we show that all necessary calculations can be performed in the Sage system. The role of periods of Abelian integrals and their commensurability in describing the nature of the winding of integral curves on an algebraic integral surface is discussed. The results of numerical experiments performed in fdm for Sage are presented.

\end{abstract}

\section{Introduction}

The idea of the motion nature of conservative dynamical systems was formed in the 1970s \cite{Moser}. A dynamical system is called transitive if almost any of its orbits fills the entire phase space densely. Transitivity of the system is prevented by conservation laws that hold the orbit on some integral manifold. Therefore, the study of a conservative dynamical system splits into two parts: 1) finding all conservation laws  by analytical methods; 2) studying a dynamical system whose restriction to integral manifolds is transitive by statistical methods.

Let us assume that the dynamic system is described by a system of differential equations 
\begin{equation}
\label{eq:ode}
\frac{d\vec x}{dt} = \vec f (\vec x),
\end{equation}
where $\vec x\in \mathbb{R}^n$, and let it possess  $r$ conservation laws
\begin{equation}
\label{eq:V}
h_i(\vec x)=c_i, \quad i=1,\dots, r,
\end{equation}
where $c_1, \dots, c_r$ are integration constants. 
The orbit $O$, or the trajectory of the system motion in the phase space $\mathbb{R}^n$, is parametrically respresentable as  
\[
\vec x = \vec x(t), \quad t \in \mathbb{R}.
\]
However, this does not mean that the orbit $O$ is a line, i.e. a manifold of dimension 1, or even that $O$  is a closed set. In fact, the closure of the orbit $O$ in $\mathbb{R}^n$ may be greater than $O$, but it cannot go beyond the integral manifold $V$ defined by the equations \eqref{eq:V}. A dynamical system \eqref{eq:ode} is called transitive on the manifold $V$ if for almost any orbit $O \subset V$ we have $\overline{O}=V$. If this equality is satisfied, then all conservation laws have been found.

The justification for the existence of a set of conservation laws \eqref{eq:V} that is complete in the sense that the dynamical system is transitive, or even better, ergodic on the manifold \eqref{eq:V}, essentially depends on the choice of the class to which the functions $h_i$ belong. Back in the 1880s, Bruns \cite{Bruns,Whittaker} developed a method for finding all integrals of motion in the class of algebraic functions for the many-body problem. This method can be extended to some Hamiltonian systems \cite{Painleve-2} and to the dynamics of a top  \cite{Kochina,Kozlov-1975}. However, this class is too narrow to single out a manifold on which the system will be transitive. For example, the predator-prey system has a transcendental integral that prevents the orbit of the system from filling the entire phase plane. Another class considered in the 20th century was the class of integrals that are meromorphic in a certain sense; this formulation goes back to the paper by Poincaré about Bruns' theorem \cite{Kozlov-1983}.

An  issue of not less importance is the development of algorithms for finding manifolds on which the system is transitive. In fact, the problem is to find a complete set of conservation laws. Note that these manifolds can be seen by observing the trajectory over a long period of time. For example, in the framework of numerical experiments with reversible dynamic systems, we have repeatedly observed how the points of the orbit found by the reversible difference scheme arrange along lines or surfaces \cite{malykh-2024-mdpi}. Thus, it is not difficult to roughly estimate the dimension of the manifold on which the integral curve ``winds'' and hence determine the number of independent conservation laws. However, the accumulation of errors in numerical methods makes conclusions based on these experiments not entirely convincing.  

A natural first step in verifying such judgments is to apply numerical methods to dynamic systems that have been exactly integrated. However, one difficulty arises here: these systems have been integrated because their trajectories are very simple. For example, nonlinear oscillators that are integrated in elliptic functions have elliptic curves as trajectories, so the question of describing the integral manifold is solved immediately, namely,  this is the well-known algebraic curve.

Interesting and non-trivial from this point of view are nonlinear dynamic systems integrable in Abelian functions of two variables \cite{Golubev-1953,Markushevich}. In this paper, we  consider the Vanhaecke system \cite{Vanhaecke-1994,Vanhaecke-2001}. This is a completely integrable Hamiltonian system with two degrees of freedom, the general solution of which is expressed by means of Abelian functions of two variables. The algebraic integral manifold of this problem is a surface. Numerical experiments show that integral curves are wound on this surface (Section \ref{n:alg}). On the other hand, from theoretical considerations it seems that the system should have integral manifolds of dimension 1, since the system is integrable both in the Liouville sense and in the sense of its solvability in classical transcendental functions \cite{Hitchin}. To describe the behavior of orbits on a two-dimensional integral algebraic manifold, we, relying on \cite{Vanhaecke-1994}, obtained the main formulas describing the behavior of the Vanhaecke system in the Sage computer algebra system and analyzed the orbits based on known results in the theory of Abelian functions \cite{Markushevich, Vanhaecke-2001, Weierstrass}.

\section{Vanhaecke system and its integral manifolds}
\label{n:alg}

The Vanhaecke system is a Hamiltonian system with two degrees of freedom and Hamiltonian
\begin{equation}
\label{eq:hamiltonian}
H=\frac{p_1^2+p_2^2}{2} + (q_1^2+q_2^2)^2 + aq_1^2 + bq_2^2
\end{equation}
for $a\not =b$, integrated by P. Vanhaecke in 1994 \cite{Vanhaecke-1994}. The trivial case $a=b$  is  briefly considered in \cite{Vanhaecke-2001}.

The Vanhaecke system has two polynomial integrals:
\[
F={\left(p_{2} q_{1} - p_{1} q_{2}\right)}^{2} - {\left(2 \, q_{1}^{4} + 2 \, q_{1}^{2} q_{2}^{2} + 2 \, a q_{1}^{2} + p_{1}^{2}\right)} {\left(a - b\right)}
\]
and
\[
G={\left(p_{2} q_{1} - p_{1} q_{2}\right)}^{2} + {\left(2 \, q_{1}^{2} q_{2}^{2} + 2 \, q_{2}^{4} + 2 \, b q_{2}^{2} + p_{2}^{2}\right)} {\left(a - b\right)},
\]
Energy does not give a new integral, since
$$
F-G=2(b-a)H.
$$
It is interesting that for $a=b$ both integrals coincide and are the square of the area integral $p_{2} q_{1} - p_{1} q_{2}$.

Equations
\begin{equation}
\label{eq:FG}
F(p_1,p_2,q_1,q_2)=f, \quad  G(p_1,p_2,q_1,q_2)=g
\end{equation}
define in $\mathbb{R}^4$ a surface $S$, the integral surface of the Vanhaecke system. We calculated its projection into $q_1 q_2 p_1$ using the Gröbner basis, it is given by the formula
\begin{equation}
\label{eq:p1}
s_2 p_1^4 + s_1 p_1^2+s_0 =0,
\end{equation}
where $s_0,s_1,s_2$ are polynomials in $q_1,q_2$, whose coefficients depend on the parameters $a,b,f,g$. For what follows, it is important to note that these polynomials contain only even powers of $q_1$ and $q_2$. Therefore, the surface \eqref{eq:p1} is symmetric with respect to reflection in the coordinate planes.

\begin{figure} 
    \centering % центрирование изображения
    \includegraphics[width=0.45\textwidth]{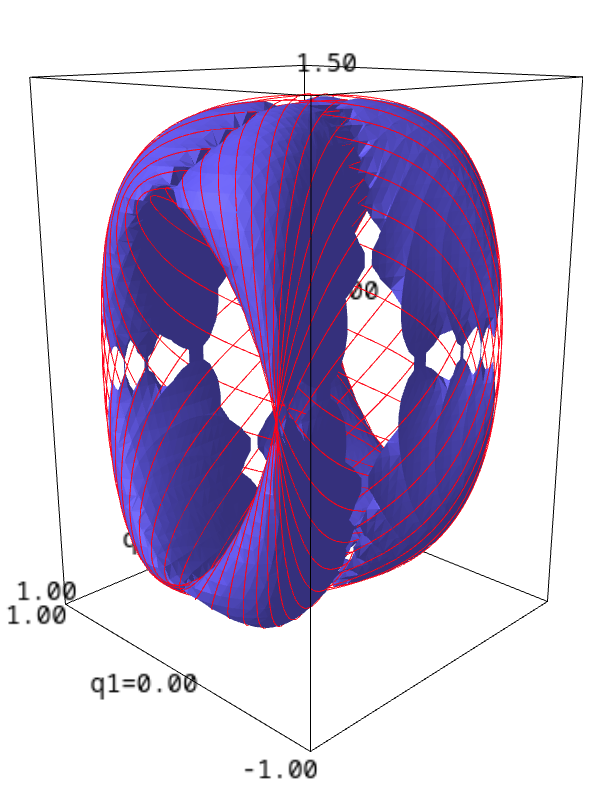} 
    \caption{Projection of the integral manifold into the space $q_1 q_2 p_1$ (blue surface) and one of the integral curves (red line) that winds onto this manifold}
    \label{fig:qp}
\end{figure}

Let us see how the integral curve is located on the integral surface. For the numerical experiments given below, we took without any intention the parameters
\begin{equation}
\label{eq:params}
a=1, \quad b=2
\end{equation}
and the particular solution at
\begin{equation}
\label{eq:initial_conditions}
p_{1} (0)= 1 , \quad p_{2} (0)= 0 , \quad q_{1} (0)= 0 , \quad q_{2} (0)= 1.
\end{equation}
We calculated the solution of the Vanhaecke system for these initial conditions using the Runge-Kutta method for $0<t<200$ with a very small step, such that its further decrease  did not lead to any noticeable change in the figures discussed below.

Fig. \ref{fig:qp} shows the projection of the integral surface \eqref{eq:p1} into the space $q_1 q_2 p_1$ and the integral curve lying on this surface. The integral curve does indeed lie on this surface and covers it densely enough for us to see the ``skeleton'' of the integral surface, but not enough to say that the integral curve winds onto this surface everywhere densely. At the same time,  the integral curve seems to be closed and therefore the solution is expected to be periodic.

Let us now discuss how the integral manifold and its projection are structured. First of all, we note that for $fg\not =0$ the integral surface has no singular points. This is easy to verify if we calculate the Gröbner basis of the ideal
\[
(F-f,G-g,F_{q_1}, \dots, G_{p_2}).
\]
The orbits of the dynamical system lie on this manifold, and they cannot intersect or have self-intersection points by virtue of the Cauchy theorem. However, when projecting from four-dimensional space to three-dimensional space, the integral surface acquires double lines on which the orbit has self-intersection points. To find the double lines of the surface \eqref{eq:p1} we calculated the Gröbner basis of the corresponding ideal and verified that there are three such lines --- these are sections of the surface by coordinate planes. For example,
\[
\left . s_2 p_1^4 + s_1 p_1^2+s_0 \right|_{q_1=0} = (p_1^2 q_2^2 - ap_1^2 + bp_1^2 - f)^2 (a - b)^2.
\]

To plot the surface in Fig. \ref{fig:qp} we used the standard function \verb|implicit_plot3d|, which cannot correctly depict the behavior of the surface near double lines. Therefore, in the figure there are quite noticeable holes along the sections by the coordinate planes. In reality, the surface in each of the octants can be described as an almost rectangular piece of the plane, bent along one of the sides and glued at three ends like dough in a varenik. A double line appears at the gluing point. Inside the piece of plane folded in this way there is a small cavity. The 8 ``vareniks'' obtained in this way are glued along double lines into a cylinder, on the surface of which the orbit lies. In Fig. \ref{fig:qp} you can see how the orbit from the outer surface of the cylinder passes through a double line to the inner side of the tube and returns back. This, in fact, prevents us from describing the situation as the winding of the orbit onto the surface. \begin{figure}
\centering % centering the image
\includegraphics[width=0.5\textwidth]{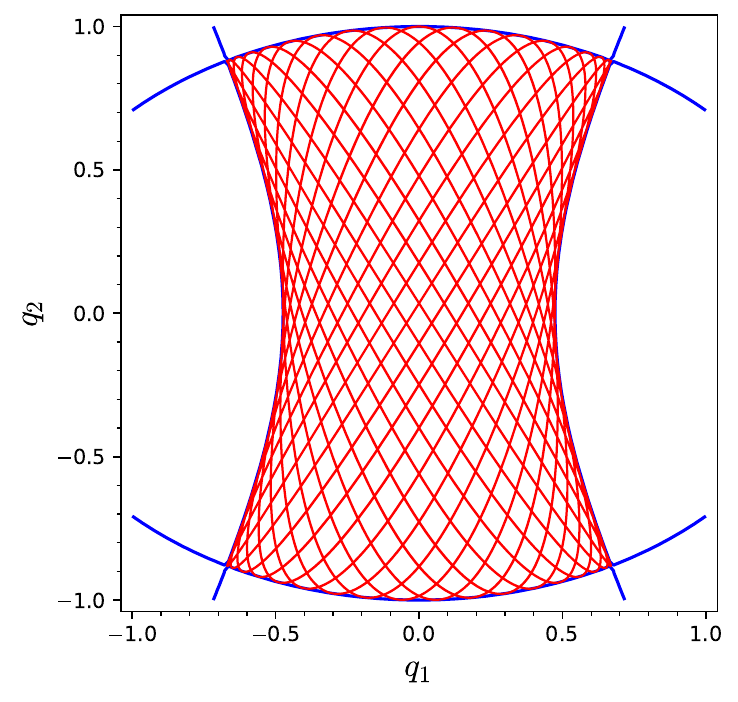}
\caption{Projection of the integral manifold onto the plane $q_1 q_2$ and one of the integral curves (red line) that winds onto this manifold}
\label{fig:q}
\end{figure}

Figure \ref{fig:q} shows the projection of this integral manifold onto the plane $q_1 q_2$. The quadratic equation \eqref{eq:p1} with respect to $p_1^2$ has real roots if its discriminant
\[
D=s_1^2-4s_2s_0 \geq 0.
\]
Therefore, the line $D=0$ on the plane $q_1q_2$ defines the boundaries of the projection of the integral surface onto this plane. This allows  depicting the boundaries of the projection of the surface $S$ in Fig. \ref{fig:q} as the zeros of the discriminant.

Looking at these illustrations, it is very difficult to answer the question about the dimension of the integral manifold on which the integral curves are wound. Much depends not on the experiment with the numerical solution itself, but on the interpreter. One can cling to the fact that the solution seems periodic and then the integral curve in Fig. \ref{fig:qp} is closed, and the dimension of the integral manifold is 1. On the contrary, one can assume that the solution is not periodic and over time the integral curve everywhere densely fills the entire integral surface. In this case, the dimension of the integral manifold is 2 and cannot be reduced.

As noted above, in the theory we are interested in the dimension of the orbit closure  as a manifold in $\mathbb{R}^4$ as $t \in \mathbb{R}$ changes. The trouble with this definition and similar-type ones is that it is not constructive. We can observe the integral curve only on a finite segment $0<t<T$ rather than on the entire interval $0<t<\infty$. At the same time, increasing $T$ leads to an increase in the error. We see in Fig.~\ref{fig:qp} that for $0<t<200$ the trajectory is a closed curve. If we increase the segment several times, the figure will not change, but the integral curve will become thicker. This can be explained both by the accumulation of rounding error and by the fact that the solution is slightly non-periodic and therefore over time the orbit will fill the entire surface.

\section{A note on Liouville's theorem}

One might think that Liouville's theorem \cite{Whittaker} allows  finding two more integrals from two polynomial integrals in involution, and therefore the system has two more integrals depending on $t$. After eliminating $t$, we should obtain a one-dimensional integral manifold, not a surface. This kind of reasoning leads to the wrong result, since it suggests carrying over local results to the global case. In fact, Liouville's theorem leads to two additional integrals of motion, which are very complicated Abelian quadratures. Vanhaecke obtains this kind of integral of motion in a different way, but, as will be shown below, it does not help at all to reduce the dimension of $\overline{O}$.

\section{Integration of the Vanhaecke system in Abelian functions}

The general scheme for integrating the Hamilton equations based on Liouville's theorem \cite{Whittaker} leads to very cumbersome quadratures. Vanhaecke managed to bypass the stage of writing quadratures from Liouville's theorem by using a convenient change of variables. As a result, it was possible to describe the solution using the Jacobi problem \cite{Markushevich}, which allows an unambiguous answer to the question posed about the dimension of the minimal integral manifold.

Vanhaecke's passage to the Jacobi problem is based on a guess, which we will describe as follows: we pass from the variables $q_1, q_2$ to two new variables $x_1, x_2$, which are roots of the equation
$$
x^2 + (q_1^2+q_2^2+a+b) x + aq_2^2 + bq_1^2 + ab=0.
$$
In other words, we change variables as
\begin{equation}
\label{eq:qx}
\left\{
\begin{aligned}
& x_{1} + x_{2} = -q_{1}^{2} - q_{2}^{2} - a - b, \\
& x_{1} x_{2} = b q_{1}^{2} + a q_{2}^{2} + a b. \\
\end{aligned}
\right.
\end{equation}
Let us leave aside the discussion of the guess itself, noting only that the appearance of $q_1^2$ and $q_2^2$ here significantly simplifies the calculations, since Eq. \eqref{eq:p1} contains only even powers of $q_1$ and $q_2$.

Let us prove in the Sage system that to find $x_1, x_2$ from the differential equations describing the Vanhaecke system, we will obtain the Jacobi problem~\cite{Markushevich}:
\begin{equation}
\label{eq:Jacobi}
\left\{
\begin{aligned}
& \frac{dx_1}{\sqrt{R(x_1)}} + \frac{dx_2}{\sqrt{R(x_2)}}=0, \\
& \frac{x_1 dx_1}{\sqrt{R(x_1)}} + \frac{x_2dx_2}{\sqrt{R(x_2)}}=dt. \\
\end{aligned}
\right.
\end{equation}
where $R$ is some polynomial of degree 5, which we will write out explicitly below.

For this purpose, we express the coordinates and momenta of the system through $x_1, x_2$ and their derivatives with respect to $t$. To do this, we differentiate \eqref{eq:qx} with respect to $t$ and find the relationship between the momenta and the derivatives of $x_1, x_2$ with respect to $t$:
\[
\left\{
\begin{aligned}
& \frac{dx_{1}}{dt} + \frac{dx_{2}}{dt} = -2 \,p_{1} q_{1} - 2 \, p_{2} q_{2} \\ 
& \frac{dx_{2}}{dt} x_{1} + \frac{dx_{1}}{dt} x_{2} = 2 \, b p_{1} q_{1} + 2 \, a p_{2} q_{2}
\end{aligned}
\right.
\] 
This system is linear with respect to $p_1, p_2$, so it is not difficult to solve it explicitly:
\[
\left\{
\begin{aligned}
& p_1 = -\frac{1}{2} \frac{(a + x_2)\dot x_1 + (a + x_1)\dot x_2}{(a - b)q_1},\\
& p_2 = \frac{1}{2} \frac{(b + x_2)\dot x_1 + (b + x_1) \dot x_2}{(a - b)q_2}.
\end{aligned}
\right.
\]
Substitute these expressions into Eqs. \eqref{eq:FG} of the surface $S$ and combine these two equations with Eqs. \eqref{eq:qx}. As a result, we arrive at a system of 4 equations containing $x, \dot x$ and $q$. Eliminating $q_1, q_2$ and expressing $\dot x_2$ through $x_1, x_2$ and constants using the Gröbner basis method, we obtain two consequences of this system:
$$
(x_1 - x_2)^2 \dot x_1^2 - R(x_1)=0
$$
and similarly
$$
(x_1 - x_2)^2 \dot x_2^2 - R(x_2)=0.
$$
In both equatilies,  $R(x)$ is a polynomial of degree 5 in  $x$:
\[
R=\frac{4}{a-b} (x+ b)(x + a)(2(a-b)x^3 + 2(a^2-b^2)x^2 + 2(a^2 b  - 2ab^2+f-g)x + bf - ag).
\]

We rewrite the obtained equalities as
$$
\frac{\dot x_1^2}{R(x_1)} = \frac{1}{(x_2-x_1)^2}, \quad \frac{\dot x_2^2}{R(x_2)} = \frac{1}{(x_1-x_2)^2}
$$
and extract the root
$$
\frac{\dot x_1}{\sqrt{R(x_1)}} = \pm \frac{1}{x_1-x_2}, \quad \frac{\dot x_2}{\sqrt{R(x_2)}} = \pm \frac{1}{x_2-x_1}.
$$
Upon permutation $x_1\to x_2$, one formula must yield another, so the sign in both formulas is either plus or minus.

But then
$$
\frac{dx_1}{\sqrt{R(x_1)}} + \frac{dx_2}{\sqrt{R(x_2)}}=0
$$
and
$$
\frac{x_1 dx_1}{\sqrt{R(x_1)}} + \frac{x_2dx_2}{\sqrt{R(x_2)}}= \pm \frac{x_1 dt}{x_1-x_2} \pm \frac{x_2 dt}{x_2-x_1}= \pm dt.
$$
Since $dt>0$, the plus sign should be chosen, which yields system  \eqref{eq:Jacobi}. 

\section{Integration in Abelian functions}

In the previous Section we came to the conclusion that the problem of integrating the Vanhaecke system on the integral surface \eqref{eq:FG} can be written as a Jacobi problem \eqref{eq:Jacobi}, which is convenient to write in quadratures
\begin{equation}
\label{eq:Jacobi:2}
\left\{
\begin{aligned}
& \int \limits_{x_1^0}^{x_1} \frac{d\xi}{\sqrt{R(\xi)}} + \int \limits_{x_2^0}^{x_2}\frac{d\xi}{\sqrt{R(\xi)}}=0, \\
& \int \limits_{x_1^0}^{x_1} \frac{\xi d\xi}{\sqrt{R(\xi)}} + \int \limits_{x_2^0}^{x_2}\frac{\xi d\xi}{\sqrt{R(\xi)}}=t, \\
\end{aligned}
\right.
\end{equation}
where $t$ is the time passed between the initial position of the system $(x_1^0,x_2^0)$ and the final position $(x_1,x_2)$. 

The fundamental theorem of the theory of Abelian functions states that every rational symmetric function $s$ of quantities $x_1, x_2$ satisfying the Jacobi problem
\begin{equation}
\label{eq:Jacobi:uv}
\left\{
\begin{aligned}
& \int \limits_{x_1^0}^{x_1} \frac{d\xi}{\sqrt{R(\xi)}} + \int \limits_{x_2^0}^{x_2}\frac{d\xi}{\sqrt{R(\xi)}}=u, \\
& \int \limits_{x_1^0}^{x_1} \frac{\xi d\xi}{\sqrt{R(\xi)}} + \int \limits_{x_2^0}^{x_2}\frac{\xi d\xi}{\sqrt{R(\xi)}}=v, \\
\end{aligned}
\right.
\end{equation}
is a meromorphic function of  $u,v$:
\[
s = \Al(u,v),
\]
referred to as Abelian function. The above is true if the polynomial $R$ has no multiple roots. For parameters $a,b,f,g$ in general position, this is indeed the case. The zeros of $R$ define the periods of the Abelian function: if $e_1, e_2$ are zeros of $R$ and there are no other zeros in the interval $[e_1, e_2]$, then the numbers
\[
\omega_1= 2\int \limits_{e_1}^{e_2} \frac{d\xi}{\sqrt{R(\xi)}}, \quad
\omega_2= 2\int \limits_{e_1}^{e_2} \frac{\xi d\xi}{\sqrt{R(\xi)}}
\]
form a system of common periods of the Abelian function:
\[
\Al(u+\omega_1, v+\omega_2)=\Al(u,v).
\]
Returning to our dynamical system, we see that
\[
x_1+x_2 =\Al_1(0,t), \quad x_1x_2=\Al_2(0,t).
\]
By virtue of \eqref{eq:qx} the squares of $q_1$ and $q_2$ are expressed linearly through these functions, which allows  expressing the general solution of the Vanhaecke system in terms of two Abelian functions of two arguments.

Note that the polynomial $R$ depends on the parameters $a,b$ and $f,g$, but does not contain  two more constants $x_1^0$ and $x_2^0$. Changing these quantities contributes additive constants to the left-hand sides of the equalities \eqref{eq:Jacobi:2}. Therefore, the formulas
\begin{equation}
\label{eq:al}
x_1+x_2 =\Al_1(c_1,t+c_2), \quad x_1x_2=\Al_2(c_1,t+c_2)
\end{equation}
determine a two-parameter family of solutions of the Vanhaecke system on a fixed integral surface \eqref{eq:FG}. Let us see how these integral curves are wound onto the algebraic integral surface. 

\section{How is the integral curve of the Vanhaecke system structured?}
\label{n:jacobi}

Let the roots of the polynomial $R$ be real, denote them as $e_1, \dots$ in ascending order. Using them it is easy to compose two real systems of common periods
\[
\omega_1=2\int \limits_{e_1}^{e_2} \frac{d\xi}{\sqrt{R(\xi)}}, \quad 
\omega_2= 2\int \limits_{e_1}^{e_2} \frac{\xi d\xi}{\sqrt{R(\xi)}}
\]
and
\[
\omega_1'=2\int \limits_{e_3}^{e_4} \frac{d\xi}{\sqrt{R(\xi)}}, \quad 
\omega_2'= 2\int \limits_{e_3}^{e_4} \frac{\xi d\xi}{\sqrt{R(\xi)}}
\]
of Abelian functions \cite{Markushevich,Weierstrass}. For any $n,m\in \mathbb{Z}$, it is true that
\[
\Al_i(c_1 + n\omega_1+m\omega_1',t+c_2+n\omega_2+m\omega_2') =\Al_i(c_1,t+c_2).
\]
For example, for the considered values of constants \eqref{eq:params} and initial conditions \eqref{eq:initial_conditions}, we have
\[
R=4(2x^2 - 3)(x + 3)(x + 2)(x + 1),
\]
therefore there are two real systems of common periods
\[
\omega_1=2\int \limits_{-3}^{-2} \frac{d\xi}{\sqrt{R(\xi)}}, \quad
\omega_2= 2\int \limits_{-3}^{-2} \frac{\xi d\xi}{\sqrt{R(\xi)}}
\]
and
\[
\omega_1'=2\int \limits_{-\sqrt{3/2}}^{-1} \frac{d\xi}{\sqrt{R(\xi)}}, \quad
\omega_2'= 2\int \limits_{-\sqrt{3/2}}^{-1} \frac{\xi d\xi}{\sqrt{R(\xi)}}
\]
\begin{theorem}
The Vanhaecke system has a periodic solution if and only if the periods $\omega_1,\omega_1'$ are commensurable.
\end{theorem}

\begin{proof}
(i) If the numbers $\omega_1,\omega_1'$ are commensurable, then we can choose $n,m\in \mathbb{Z}$ so that
\[
n\omega_1+m\omega_1'=0
\]
and then $q_i^2$ have period $n\omega_2+m\omega_2'$. In this case, the integral curve is a closed curve and the dimension of the minimal integral manifold is 1. 

(ii) Let the solution have period $T$, then
\[
\Al_i(u,v+T) =\Al_i(u,v).
\]
If the numbers $\omega_1,\omega_1'$ are commensurable, then $q_i^2$ are not periodic functions. Then for some $T$ we have
\[
\Al_i(u,v+T) =\Al_i(u,v).
\]
Every pair of common periods of an Abelian function can be expressed as a linear combination of four pairs of common periods with integer coefficients of the famous theorem on the fundamental system of periods proved by Weierstrass \cite{Weierstrass}. Two of these four pairs are imaginary, so they can be omitted. But then
\[
0 = n\omega_1+m\omega_1',
\]
and so the periods are commensurable.
\end{proof}

Note that the polynomial $R$ depends on the parameters $a,b$ and $f,g$, but not on the choice of initial conditions on the integral surface \eqref{eq:FG}. Therefore, either all integral curves on a fixed surface \eqref{eq:FG} are closed, or all are open. And this depends on the relationship between the two numbers  $\omega_1$ and $\omega_1'$.

\begin{figure} 
    \centering % центрирование изображения
    \includegraphics[width=0.8\textwidth]{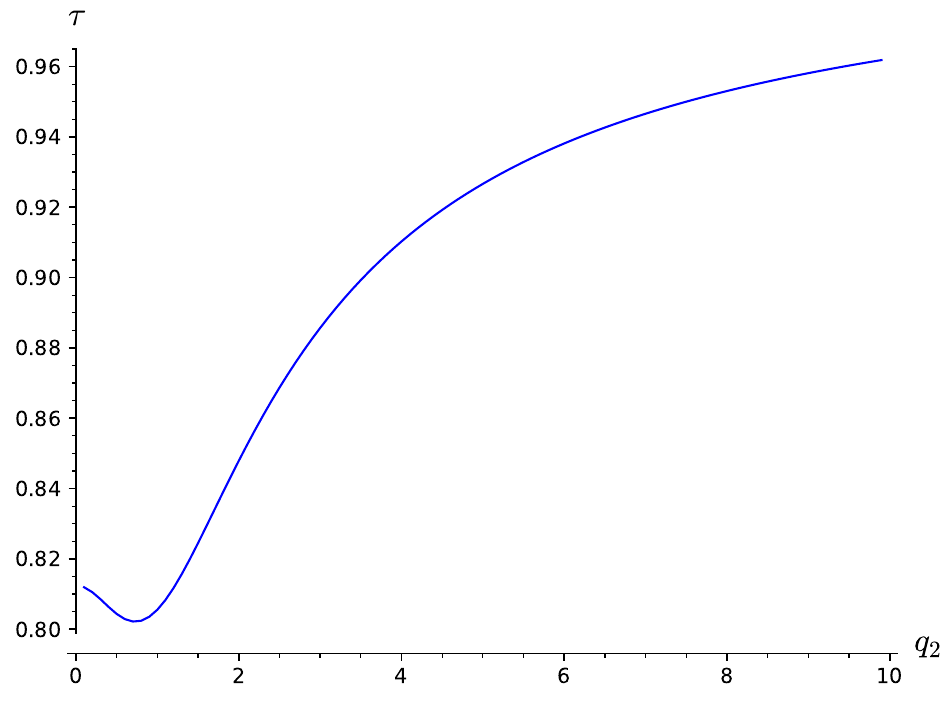} 
    \caption{Dependence of $\tau$ on the choice of the initial value of  $q_2$ for $q_1=0$, $p_1=1$, and $p_2=0$} 
    \label{fig:tau}
\end{figure}

The ratio
\[
\tau = \frac{\omega_1}{\omega_1'}
\]
 depends on parameters  $f$ and $g$ in a continious way, see Fig. \ref{fig:tau}. Therefore, by changing these parameters arbitrarily little, we can obtain either a rational number or an irrational one. Thus, in the neighborhood of any point of the four-dimensional space $q_1q_2p_1p_2$ there are infinitely many integral surfaces on which the integral curves are closed, and infinitely many surfaces on which they are not closed.

If for the considered values of the constants $f,g$ the quantity $\tau$ is rational, i.e.,
\[
\tau(f,g)=-m/n \in \mathbb{Q},
\]
then any solution of the Vanhaecke system, the initial data of which are taken on the surface \eqref{eq:FG}, is periodic, and the real number
\[
T=n\omega_2+m\omega_2'
\]
is the period of these solutions. The orbits corresponding to periodic solutions are smooth closed curves without self-intersection. Therefore, the dynamical Vanhaecke system is not transitive on the algebraic surface \eqref{eq:FG} for such values of $f,g$.

The periods of Abelian functions depend on $R$, that is, on $f,g$, but not on the choice of initial conditions on the integral surface \eqref{eq:FG}. Therefore, all solutions lying on the surface \eqref{eq:FG} for such values of $f,g$ have the same period.
This allows revealing the integral surface \eqref{eq:FG} in numerical experiments if  considering all solutions having the same period rather than one individual periodic solution. We proposed to call the manifold formed by the points of all possible solutions having the same period an equiperiodic manifold \cite{Malykh-2022-pomi}. In the case of the Vanhaecke system, equiperiodic manifolds are algebraic surfaces belonging to the family of integral surfaces \eqref{eq:FG} and correspond to those particular values of the parameters $f,g$ for which $\tau \in \mathbb{Q}$.

A small rational change in $\tau$ in the topology of $\mathbb{R}$ can lead to a huge change in the period. Thus, the value of the period is not stable to small changes in the parameters of the integral surface. This does not at all contradict the stability of the initial problem solution with respect to a change in the initial data, i.e,  a change in $f$ and $g$. If in the unperturbed state we had a solution with a period of $T=n\omega_2+m\omega_2'$, then with compensation we obtain a solution that almost returns back in time $T$.

The behavior of the integral curve for irrational $\tau$ can be deduced from the following theorem.

\begin{theorem}[Jacobi theorem on infinitesimal periods]
\label{th:Jacobi}
Let $\omega$ and $\omega'$ be two real numbers. Then either they are commensurable, i.e., there are two integers $n$ and $m$ such that
\[
n \omega + m\omega'=0,
\]
or they are incommensurable, i.e.,
\[
n \omega + m\omega'\not =0 \quad \forall n,m\in \mathbb{Z},
\]
but for any $\epsilon>0$ there are integers $n$ and $m$ such that
\[
|n \omega + m\omega'|<\epsilon
\]
\end{theorem}

In the absence of a suitable source, we present a proof. It repeats the Euclidean integer division algorithm.

\begin{proof}
Without loss of generality, we can assume that $|\omega|>|\omega'|$. Then we can choose a number $n\in \mathbb{Z}$ such that
\[
|\omega-n\omega'|\leq \frac{|\omega'|}{2}
\]
We set
\[
\omega''=\omega-n\omega', \quad |\omega''|\leq \frac{|\omega'|}{2}.
\]
But then $|\omega'|>|\omega''|$ and we can choose a number $n'\in \mathbb{Z}$ such that
\[
|\omega'-n'\omega''|\leq \frac{|\omega''|}{2}
\]
Let us put
\[
\omega'''=\omega'-n'\omega''=-n'\omega+m'\omega', \quad |\omega'''|\leq \frac{|\omega''|}{2}.
\]
Proceeding in this way, we obtain the number
\[
\omega^{(n)}= n\omega+ m\omega', \quad n,m\in \mathbb{Z},
\]
for which the estimate
\[
|\omega^{(n)}|\leq \frac{|\omega^{(n-1)}|}{2} \leq \dots \leq \frac{|\omega'|}{2^{n-1}} is valid.
\]
Thus, we obtain an infinitesimal sequence $\{\omega^{(n)}\}$, which was to be proved. 
\end{proof}

Note that for 
\[
\Delta  u = n\omega_1+m\omega_1', \Delta v = n\omega_2+m\omega_2'
\]
it is true that
\[
\Al_i(c_1+k\Delta u,t +c_2)=\Al_i(c_1, t-k\Delta v+c_2), \quad k \in \mathbb{Z}.
\]
From the Jacobi theorem \ref{th:Jacobi} it follows that $\Delta u$ can be taken arbitrarily small. Therefore   $k\Delta u$ can be take close to any specified number   $u\in \mathbb{R}$. But then the integral curve 
\[
x_1+x_2 =\Al_1(0,t)=\Al_1(k\Delta u,t-k\Delta v), \quad x_1x_2=\Al_2(0,t)=\Al_2(k\Delta u,t-k\Delta v)
\]
passes arbitrarily close in the topology of $\mathbb{R}$ to any point on the surface
\[
x_1+x_2 =\Al_1(u,v), \quad x_1x_2=\Al_2(u,v), \quad (u,v) \in \mathbb{R}^2.
\]
Therefore, the closure of the set of all points of the integral curve \eqref{eq:al} is a surface and the dimension of the integral manifold onto which the orbit winds is equal to 2.

Since the set of rational numbers is inferior in power to the set of irrational numbers, it seems surprising that we see a periodic curve in Fig. \ref{fig:qp}. Random initial data should not yield closed curves. This effect is again explained by the Jacobi theorem. Taking $n,m$ such that $n\omega_1+m\omega_1'$ is arbitrarily small in absolute value, we  see, by virtue of the uniform continuity of Abelian functions, that the equality
\[
q_i^2(t+n\omega_2+m\omega_2')\simeq q_i^2(t)
\]
is satisfied with arbitrarily high accuracy. 

\begin{figure} 
    \centering % центрирование изображения
    \includegraphics[width=0.8\textwidth]{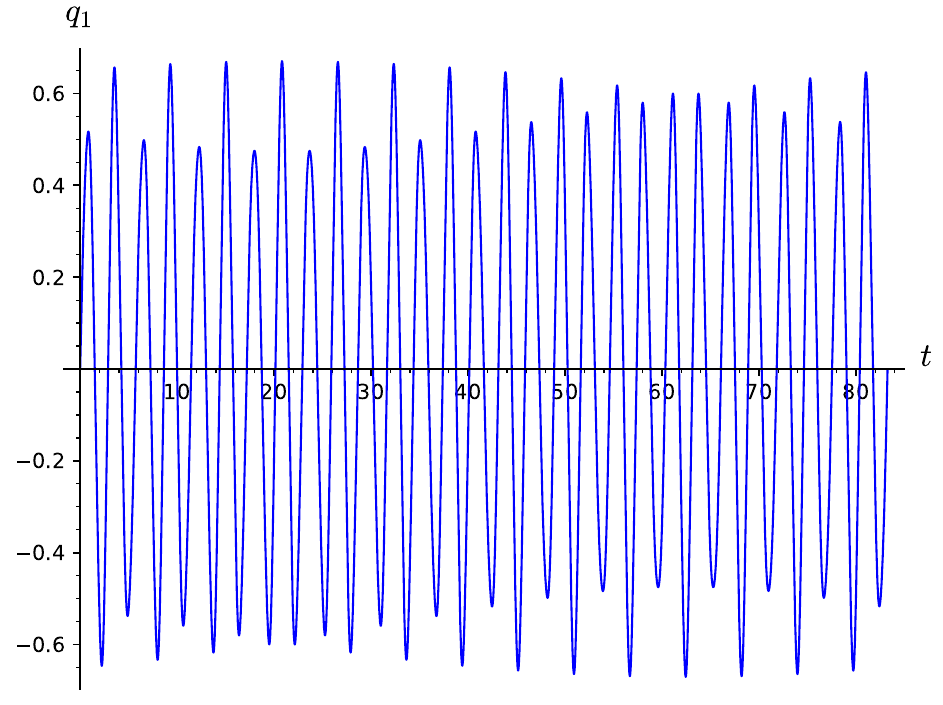} 
    \caption{Dependence of $q_1$ on $t$ in an interval equal to twice the approximate period $29 \omega_2' - 72 \omega_2$} 
    \label{fig:q1}
\end{figure}

This is exactly the case we observe in our example. The equality
\[
\frac{\omega_1}{\omega_1'}\simeq \frac{29}{72}
\]
is satisfied with very good accuracy:
\begin{equation}
\label{eq:29/72}
29 \omega_1' - 72 \omega_1 = 1 \cdot 10^{-4},
\end{equation}
i.e., two-digit $n,m$ give a 4th-order zero. Therefore, in numerical experiments it seems that the solution is periodic and the squares $q_i$ have a period
\[
29 \omega_2' - 72 \omega_2=41.624490476319636.
\]
It is easy to see that $q_i$ themselves are single-valued functions of $t$, and their period is twice as large, see Fig. \ref{fig:q1}.

\begin{figure} 
    \centering % центрирование изображения
    \includegraphics[width=0.8\textwidth]{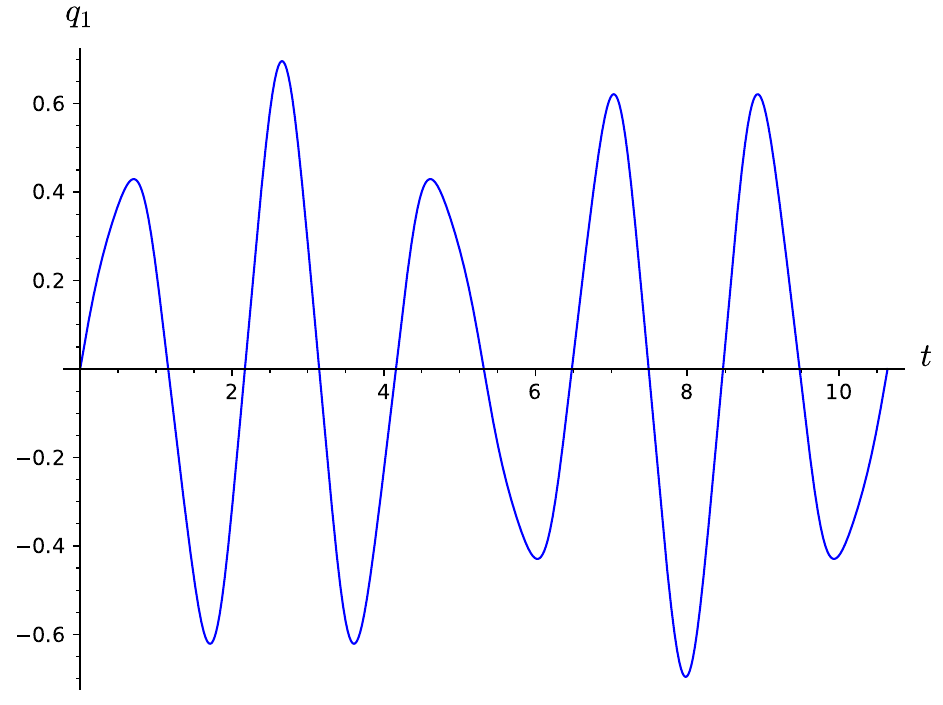} 
    \caption{Dependence of $q_1$ on $t$ in an interval equal to twice the period $-6\omega_2+5\omega_2'$} 
    \label{fig:q1:bis}
\end{figure}

\begin{figure} 
    \centering % центрирование изображения
    \includegraphics[width=0.45\textwidth]{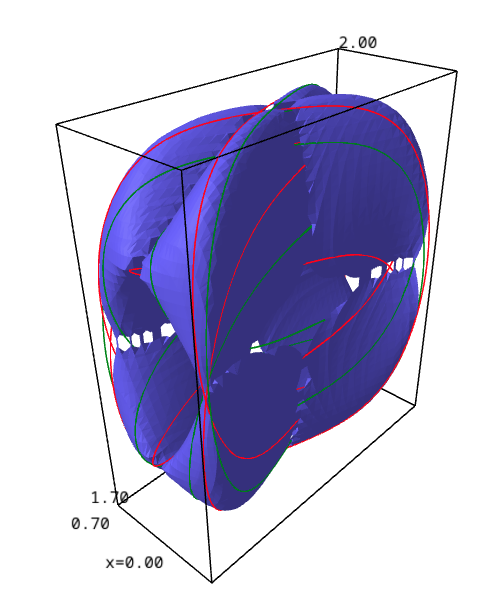} 
    \caption{Projection of the integral manifold into the space $q_1 q_2 p_1$ for initial conditions corresponding to $\tau=5/6$ and two orbits lying on this surface (red and green)} 
    \label{fig:qp:bis}
\end{figure}

The interval of $\tau$ variation in Fig. \ref{fig:tau} is not large, but it contains the value $5/6$. It is easy to select the initial values corresponding to such a value of $\tau$ from this plot:
\[
q_1=0, \quad q_2=1.688?, \quad p_1=1, \quad p_2=0.
\]
In this case, the period of oscillations for the squares $q_1$ and $q_2$ will be equal to
\[
T=-6\omega_2+5\omega_2'=5.31985?,
\]
which can be easily verified from the plot of the solution found by the Runge-Kutta method (Fig. \ref{fig:q1:bis}).

Fig. \ref{fig:qp:bis} shows the projection of the integral manifold into the space $q_1 q_2 p_1$ for this case. As noted above, it is eight bodies linked into a cylinder along double lines. But now, the cavities in these bodies are larger and therefore more noticeable. The orbit covers the surface less densely and is easier to follow. Now it is clearly visible how the trajectories pass onto the inner surface of the pipe through double lines. It is interesting to note that both trajectories shown in Fig. \ref{fig:qp:bis} have the same period, but this is not at all visible in this figure.

Numerical experiments suggest that the larger in absolute value the integers $n,m$, the ratio of which gives $\tau$, the more oscillations the periodic solution makes over the period and the more densely it covers the integral surface, and, exactly, more densely, but not everywhere densely.

\section{Conclusion}

The Vanhaecke system is a completely integrable Hamiltonian system. Its orbits can be divided into two classes, periodic and non-periodic. The points of the orbits corresponding to solutions with the same period form an equiperiodic manifold. In this case, there are infinitely many different equiperiodic manifolds, and all of them are algebraic integral surfaces belonging to the family \eqref{eq:V}.

For arbitrary initial data, the orbit is not closed and fills at $t \in\mathbb{R}$ some piece of the algebraic surface of the surface everywhere densely. In other words, for almost any initial data, the restriction of the Vanhaecke system to an integral surface is transitive. This does not contradict either the complete integrability of the Vanhaecke system in the sense of Liouville or its solvability in classical transcendental functions. The very intricate course of the orbit on the surface is described analytically by means of Abelian functions of two variables and is determined by four real numbers, a set of periods of Abelian functions. The key point here is the analytical description of functions of one variable $t$ using periodic functions of two variables.

The behavior of orbits in numerical experiments is determined not so much by the rationality or irrationality of the ratio of two periods $\tau$, but by the possibility of approximating this number with a rational fraction with high accuracy. Therefore, for arbitrary initial data, we see a seemingly periodic solution with a very large period and a closed orbit that runs along the integral surface densely enough for this surface to become visible, but not densely enough to allow saying that it covers the surface everywhere densely.

We believe that the Vanhaecke system is a good test  to hone methods for studying integral manifolds of conservative dynamical systems.

\begin{acknowledgments}
The work was carried out with the financial support of the Russian Science Foundation (project No. 20-11-20257). 
\end{acknowledgments}

\bibliographystyle{unsrt}
\bibliography{abh}
  
\end{document}